\documentclass[a4paper, oneside, twocolumn, notitlepage, 10pt]{extarticle_ecoc}
\usepackage{ecoc}
\usepackage{comment}
\usepackage{svg}

\begin{document}
\selectlanguage{english}    


\title{Global Seismic Monitoring using Operational Subsea Cable}%


\vspace{-0.8cm}
\author{
    M.~Mazur\textsuperscript{(1)}, 
    M.~Karrenbach\textsuperscript{(2)}, 
    N.~K.~Fontaine\textsuperscript{(1)},      
    R.~Ryf\textsuperscript{(1)}, 
    V.~Kamalov\textsuperscript{(3)}, \\ 
    L.~Dallachiesa\textsuperscript{(1)},
    \"O.~Jonsson\textsuperscript{(4)}, 
    A.~Arnar~Hlynsson\textsuperscript{(4)}, 
    S.~Hlynsson\textsuperscript{(4)},
    H.~Chen\textsuperscript{(1)},
    D.~Winter\textsuperscript{(1)}, \\
    D.~T.~Neilson\textsuperscript{(1)}, 
    A.~Ruiz-Angulo\textsuperscript{(5)} and 
    V.~Hjorleifsdottir\textsuperscript{(6)}  
}

\maketitle                  
\begin{strip}
\begin{author_descr}

\textsuperscript{(1)} Nokia Bell Labs, 600 Mountain Ave., Murray Hill, NJ 07974, USA \\
\textsuperscript{(2)} Seismics Unusual, LLC, Brea, CA 92821, USA \\
\textsuperscript{(3)} Valey Kamalov LLC, Gainesville FL 32607 USA\\
\textsuperscript{(4)} Farice, Gudridarstigur 2-4, 113, Reykjavik, Iceland \\
\textsuperscript{(5)} Institute of Earth Sciences, University of Iceland, Reykjavik, Iceland \\
\textsuperscript{(6)} Department of Engineering, Reykjavik University, Reykjavik, Iceland \\
 
 \end{author_descr}
 \vspace{-6mm}
\end{strip}
\setstretch{1.1}
\renewcommand\footnotemark{}
\renewcommand\footnoterule{}

\vspace{-1mm}
\begin{strip}
  \begin{ecoc_abstract}
We report tele-seismic waves detection from multiple earthquakes on an operational subsea cable from Iceland to Ireland. Using per-span laser interferometry with 100\,km spacing, we report clear detection of S-, P- and surface waves from multiple world-wide earthquakes, enabling seismic analysis for early warning applications.
 ~\textcopyright2024 The Author(s) \textcolor{blue}{\uline{mikael.mazur@nokia-bell-labs.com}}
 \end{ecoc_abstract}
\vspace{-3mm}
\end{strip}

\section{Introduction} \vspace{-2mm}
\label{sec:intro}
The growing interest in using the global fiber network for environmental sensing is driven by the need for monitoring in sparsely instrumented areas, such as the deep ocean. Despite covering 71\% of the Earth's surface, the deep ocean remains largely unmonitored and Earth and climate sciences could significantly benefit from enhanced ocean monitoring.
The  Global Ocean Observing System (GOOS)\cite{GOOS}, which is led by UNESCO, focuses on three key areas: climate, forecasts and warnings and ocean health. 
Ocean monitoring for forecasting and warnings requires real-time data from globally distributed sensors~\cite{GOOS}. This poses challenges for deep ocean sensors due to the need for continuous power and data connection. Current monitoring arrays, like Donet II and S-Net~\cite{DONET} off Japan and the five arrays of the Ocean Observatories Initiative \cite{OOI} deployed off the coast of North America and Greenland, are primarily located near coastlines, leaving vast areas of the ocean unmonitored.
\begin{figure*}[ht]
   \centering
    \includegraphics[width=1\linewidth]{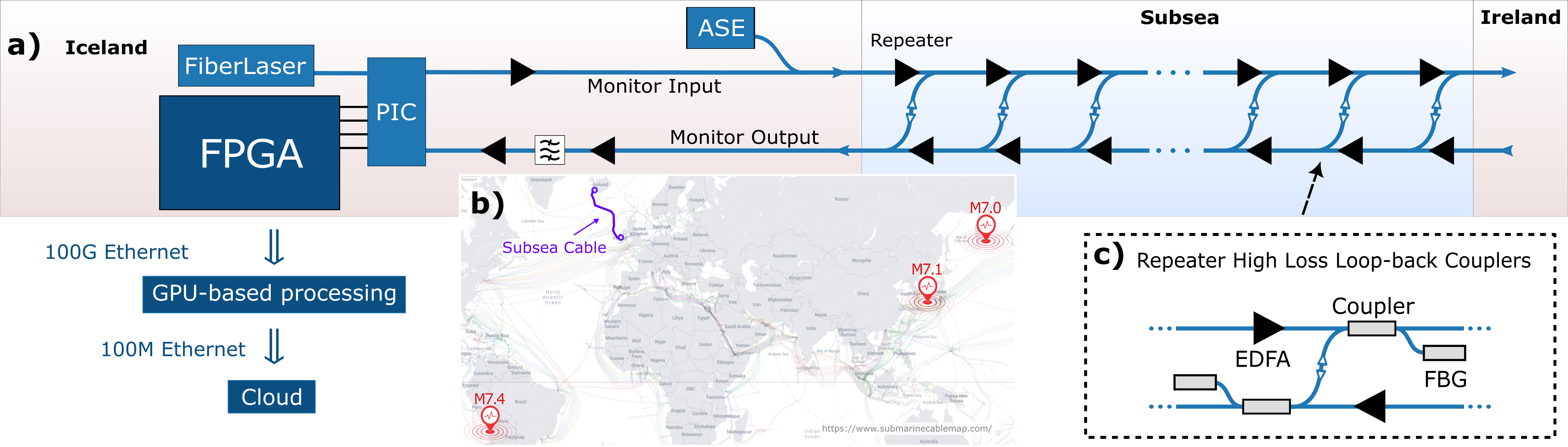}
    \caption{ (a) Field deployed multi-span OFDR sensing system, used for seismic monitoring. System consist of a silicon photonic integrated circuit, an FPGA and GPU connected via high-speed Ethernet for real-time processing and receiver with a narrow ($<$100\,Hz) linewidth laser. (b) Location of submarine fiber and earthquakes. The sensing fiber is a multi span amplified subsea cable system running from Iceland to Ireland, with (c) loop back couplers with Bragg reflectors. }
    \label{fig:setup}\vspace{-1mm}
\end{figure*}

Early demonstrations of using operational telecom cables for seismic monitoring has yielded promising result\cite{marra2022optical,mazur2023advanceddistributedsubmarinecable,costa2023localization,Mazur2024cont}. Compared to standard fiber sensing implementations using distributed acoustic sensing (DAS), which is limited to the first/last span, this enables distributed measurements along the entire subsea cable. Especially for early warnings such as tsunami and earthquakes, have sensors far off shore is key. In the deep ocean, Tsunami waves travel at 800\,km/hr but the amplitude is tiny. To predict the impact of an earthquake, and decide weather or not to issue a tsunami alarm, details of the seismic event must be measured~\cite{mori2022giant}. This requires the detection using multiple seismic stations capturing the arrival of several wavefronts. While previous demonstrations have shown detection of earthquakes on multiple spans, the observations have been limited to surface waves, which are the largest magnitude waves typically responsible for large destruction. However, these waves are presided by so called P- and S-waves, which rather than travelling along the Earth's surface, represent the two polarizations of the seismic waves travelling the shortest path through the Earth's interior, at speeds of around 6\,km/s~\cite{zhao1992tomographic}.

Here we report detection of S-, P- and surface waves from three large magnitude earthquakes on a Atlantic subsea cable from Iceland to Ireland. We benchmark our observed waveforms to land-based reference station in Iceland~\cite{BORG}, showing a good agreement between the two. Detailed analysis of the waveforms reveal a slight difference in arrival times between the different segments of the cable and the land-based stations.  During 3 months of continuous measurements, about 20 earthquakes were detected on several spans of the subsea cable. We focus our detailed analysis on the August 8th, magnitude 7.1 earthquake in Japan noting that this earthquake caused a small coastal tsunami (about 0.4m) as well as triggered the large-scale evacuation alarms in Japan \cite{Hyuganada}. We use per-span laser interferometry to retrieve the waveforms, showing clear detection of all 17 spans. The ability to resolve several wavefronts from the earthquake demonstrate that useful seismic information can be extracted from operational subsea cables to complement the gap between land-based stations. Our results demonstrate the feasibility of including operational telecom cables into global ocean hazard monitoring systems, with the potential to add 25,000 seismic stations, as well as using data collected from remote areas to improve velocity tomographic models of the Earth.  

\begin{figure*}[ht!]
   \centering
    \includegraphics[width=.95\linewidth]{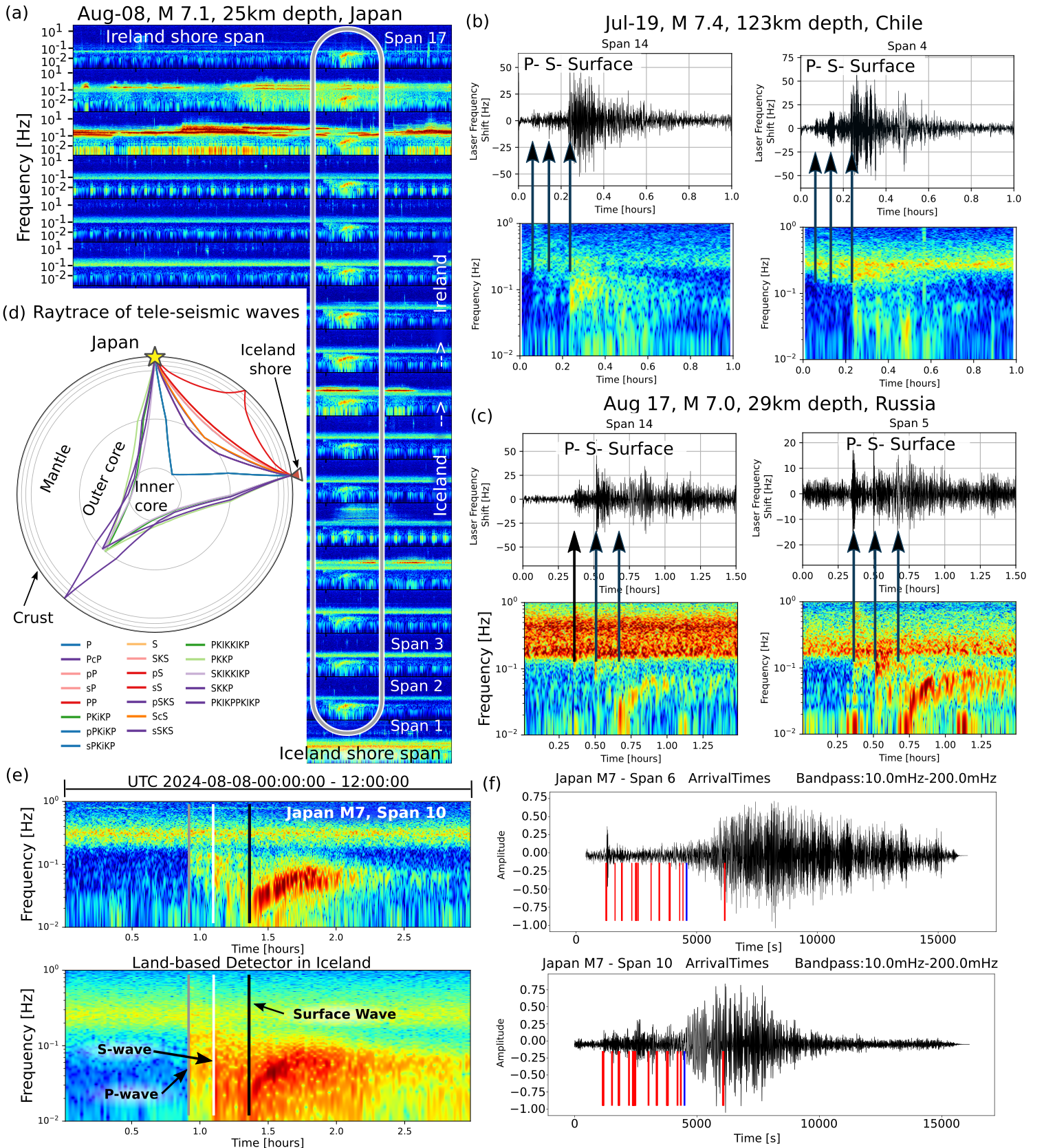}
    \caption{(a) Spectrograms from all 17 spans from Aug 8th 2024 UTC00:00:00 to UTC12:00:00 from the submarine cable monitoring system showing the recorded Japan earthquake~\cite{Jap080824,}  (highlighted). (b) and (c) Selected time-domain waveforms and spectrograms for the July 19th\cite{Chi071924} and Aug 17th\cite{Rus081724} earthquakes, respectively. Vertical scale is the measured laser frequency shift which is proportional to the seismic amplitude. (d) Raytrace diagram of tele-seismic wave propagation from Japan to Iceland. (e) Spectrograms comparing Span 9 of the subsea cable with a broadband seismometer placed in a borehole in Iceland~\cite{BORG}. (f) Zoom in comparing spans 6 and 10 with the arrival times computed from (d). (f) Detailed observation reveals that several of the non-direct paths can be identified. }
    \label{fig:res}\vspace{-1mm}
\end{figure*}


\vspace{-3mm}
\section{Experimental Setup}\label{sec:exp}
Fig.~\ref{fig:setup} shows the cable monitoring system consists of a photonic integrated circuit (PIC), an FPGA and a streaming-capable GPU and an NKT BASIC X15 laser. It is an improved version from~\cite{mazur2023advanceddistributedsubmarinecable} utilizing improved real-time processing and PIC technology to create a scalable interrogator solution. Chirped pulses with 250\,MHz bandwidth were used.  
The FPGA+GPU combination performed the real-time processing using standard OFDR-based techniques~\cite{Moore2011}, extracting the complex Jones matrix and phase of each repeater. 
The IRIS cable is about 1770\,km long connecting Iceland to Ireland, and operated by Farice. The cable sensing system is placed at the cable landing station in Iceland. The sensing wavelength was tuned to overlap with the specified repeater monitoring wavelength, exploiting the fiber Bragg gratings (FBGs) built into each repeater to enable remote monitoring. Fig.~\ref{fig:setup}(c) presents the repeater configuration and the high-loss loopback.\vspace{-2mm}


\vspace{-1mm}
\section{Results} 
Measured spectrograms from all 17 spans from UTC00:00:00 to UTC12:00:00 August 8th are shown in Fig.~\ref{fig:res}(a). 
Overall, during the 3 months of measurement time 15 earthquakes were detected on all spans, with several more being only detected on a few spans. Figure ~\ref{fig:res}(b) and (c) show time-domain waveforms and spectrograms from two selected spans for two additional large-scale earthquakes. Importantly, while the earthquake epicenters \cite{Chi071924,Jap080824,Rus081724} shown in Fig.~\ref{fig:setup}(b) are in South America, Japan and East Russia, the cable records distinct signatures on all 17 spans. In addition to the earthquakes, several other movements can be observed. We note that span 15-16 are active, showing strong oscillations around -Hz level. These spans correspond to the cable leaving the deep ocean and entering shallower water of the northern coast of Ireland and these movements are likely due to ocean currents and swells affecting the cable in areas with weak seismic coupling, leading to large-scale oscillations. 
Span 0, formed using a reflector  within the landing station, exhibits higher noise levels due to its mixed land and shallow water environment. While interesting, it's within the reach of traditional DAS systems, so we focus on the other spans.

Seismic waves, including primary (P-), secondary (S-), and surface waves, arrive at the cable at distinct times that can be computed based on Earth's internal velocity models \cite{iasp91}. The P-wave, arriving first after 12.3 minutes, is compressive. The S-wave, arriving 10 minutes later (22.5 minutes total), oscillates perpendicularly. Finally, the surface wave, easily identifiable due to its dispersive nature, arrives after 43.5 minutes.
A comparison between the recorded spectrogram from span 10 of the subsea cable and a borehole seismic station in Iceland~\cite{BORG} are shown in Fig.~\ref{fig:res}(d). We note that the qualitative agreement is good, and again both the P-, S- and surface waves can easily be identified. Importantly, some difference is attributed to directionality, with the seismic instrument used being aligned to the Z-axis with respect to ground. Given that the submarine cable only provides a single axis, accurate calibration to account for directional dependence is needed for performing absolute measurements. 
Finally, focusing on using operational subsea cables for earthquake detection and early warning applications we note that in practise, several additional waves originates from a single earthquake event. The resulting raytrace diagram for the P- and S- waves is shown in Fig.~\ref{fig:res}(d) with listed phases~\cite{storchak2003iaspei}. We create 13 distinct groups, each with waves arriving within the same minute, with the first one being the P-waves and the last one the direct surface waves. A direct comparison for span 4 and span 10 are shown in Fig.~\ref{fig:res}(e) and (f) respectively. Zooming in, we observe that several of the additional phases, such as the PP and PK phases (containing the PKiKP, pPKiKP and SPKiKP waves) arriving after 15 and 18 minutes, respectively, also can be observed on multiple spans.

In conclusion, this work demonstrates the feasibility of using operational subsea telecommunication cables for seismic monitoring, showcasing the detection of S-, P-, and surface waves from multiple earthquakes. 
The analysis of waveforms from a 17 spans forming the 1770 km cable connecting Iceland to Ireland reveals clear signatures of seismic events, even those originating from distant locations. 
This highlights the potential of subsea cables to provide distributed sensing which can complement land-based seismic networks, particularly in the vast and under-monitored deep ocean. 




\clearpage

\printbibliography
\end{document}